Nonstructural acousto-injection luminescence in metalized lithium niobate wafers.

Igor Ostrovskii [a], Oleg Korotchenkov [b], Nikolaj Borovoy [b], Andriy Nadtochiy [a,b], Roman Chupryna [b], Chandrima Chatterjee [a]

.

[a] University of Mississippi, Department of Physics and Astronomy, Oxford, MS 38677, U.S.A.

[b] Taras Shevchenko National University of Kyiv, Kyiv 01601, Ukraine.

We report the observations of nonstructural acousto-injection luminescence (NAIL) from metallized $LiNbO_3$ wafers. The samples under study are the X- or Y-cut wafers with the silver paste electrodes on opposite surfaces. The thickness is 1 – 2 mm and other linear dimensions are in the range of 2 to 25 mm. Experiments are done at room temperature in the MHz-frequency range. Mainly the fundamental shear ultrasound modes are excited in the samples. Experimentally are measured the spectra of NAIL, photo-luminescence spectra, acousto-electric properties, X-Ray diffraction under NAIL. The NAIL and other effects appear above certain threshold amplitude of the acoustical strain that is about $10^{-5}$. The involvement of the microstructural non-uniformities in the effects observed is experimentally identified by the photo-luminescence of point defects taken from different micro-regions of the samples. The data are explained in terms of the considerable acoustic stresses and piezoelectric fields that are capable injecting charge carriers from the metal contacts into a crystal.

## 1. Introduction

The Lithium Niobate ferroelectric crystal $LiNbO_3$ (LN) is very important for fundamental research and applications [1-3]. Traditionally useful physical properties of LN such as strong piezoelectric effect, radiation hardness and efficient acousto-optic constants allow fabricating several high-technology elements for telecommunications and Ultrasonics [4, 5]. For instance, there are ultrasonic transducers for medical imaging and ultrasonic, radio-frequency filters that are installed in modern cell phones [4]. That is why new observations from LN may have a potential for better understanding basic phenomena of charge transfer and energy exchange in LN along with possible novel applications.

The effect of acousto-luminescence (AL) consists in energy transformation from ultrasound into light in solids [6 - 9]. It was demonstrated three different types of AL in crystals. They are connected to the following physical processes: 1) The acoustically stimulated redistribution of electric charge carriers involving recharging of the donors and acceptors with the subsequent radiative transitions from these levels within a band gap of a crystal [6]; 2) The type of acousto-electro-luminescence when a piezo-electrically active wave generates an electric field that penetrates into an adjacent media outside the crystal, and then a luminescence is excited in the adjacent media [7, 8]; 3) Acousto-electric injection of charge carriers from a metal contact into a piezoelectric crystal such as CdS or $LiNbO_3$ with a following radiative recombination of the injected carriers through the local energy levels within a band gap [9]. The acousto-electric injection itself occurs due to piezoelectric field generated by ultrasound. The subsequent recombination of injected charges through the crystal defects having distinct energy levels inside a band gap contains the optical peaks associated with these defects. This last luminescence may be named as structural acousto-injection luminescence (SAIL). All known types of AL have the bands or lines in their spectra. The spectrum of SAIL taken under respectively high level of excitation by rf-voltage (U) is exemplified in Fig.1 taken from reference [9]. The difference of the two spectra (1 and 2) is given by plot 3, which represents a contribution of radiative transitions in a system of intra-band energy levels due to point defects in LN crystal lattice.

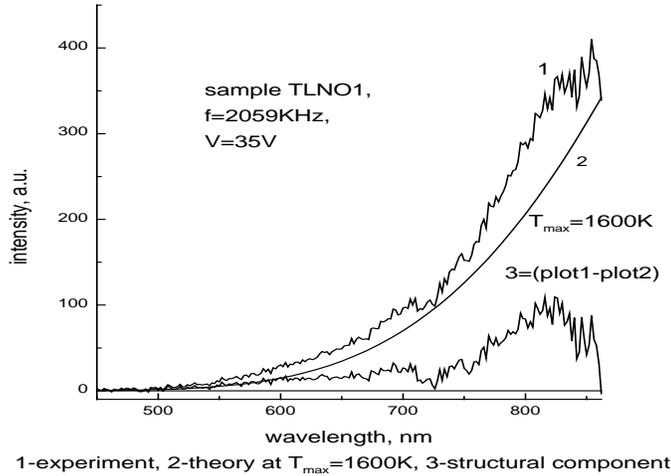

1-experiment, 2-theory at $T_{max}$=1600K, 3-structural component

Fig. 1. Spectrum of SAIL from metalized $LiNbO_3$ sample TLNO-1 at U = 35 V. Plot 1 – detected spectrum, plot 2 depicts the best fit of the nonstructural component to blackbody radiation assuming the temperature in the emission region of 1600 K, plot 3 – difference of spectra 1 and 2.

The physical origin of the nonstructural component given by plot 2 in Fig. 1 is not known yet. One can try to interpret it as being due to blackbody radiation. However, there is the apparent discrepancy between the measured spectrum (1 in Fig. 1) and that of a blackbody emitter (spectrum 2). Fits to a blackbody spectrum can be attempted for different temperatures, with best results for about 1600 K (spectrum 2 in Fig. 1). At temperatures greater than 1600 K, the Planck intensity remains a bit closer to the experimentally measured SAIL intensity in the spectral range from 500 to 675 nm, but the discrepancy of the two is even more pronounced for greater wavelengths.

Another physical problem of approximating the SAIL spectra by blackbody radiation is the fact that the melting point of LiNbO$_3$ is 1530 K, which is smaller than that required for the blackbody fit. Therefore, a model of pure blackbody emission cannot account for the experiments given in Fig. 1.

Present work is devoted to systematic investigations of the above broad nonstructural acousto-injection luminescence (NAIL) in metalized lithium niobate (MLNO). In order to emphasize the likely mechanism that is consistent with the NAIL experimental data, we also provide a discussion of the electro-acoustic and photoluminescence (PL) measurements from MLNO plates emitting NAIL.

## 2. Experimental

The experimental setup is sketched in Fig. 2. It is important to note that, in order to decrease the structural component in AL (spectrum 1 in Fig. 1), the exciting rf-voltage amplitude $U$ has to be kept rather low, typically below $U \approx 30$ V in our experimental configuration (see Fig. 2). The experiments are conducted without any external illumination or heating of the samples.

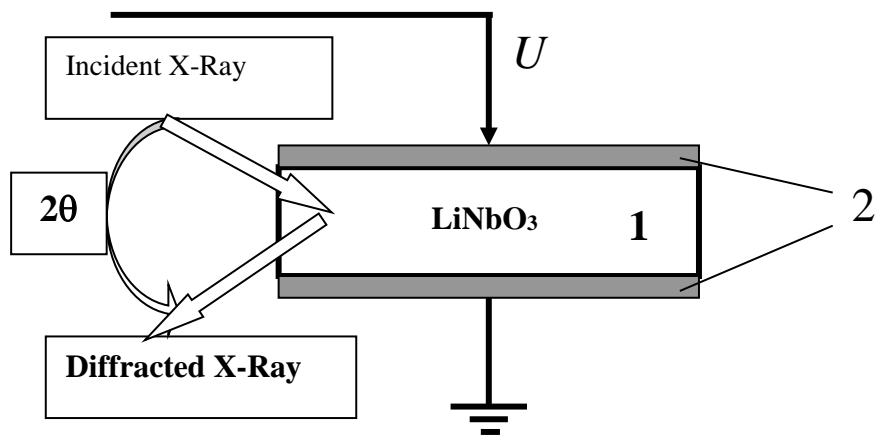

Fig. 2. Experimental setup for X-Ray measurements: 1– LiNbO$_3$ plate, 2 – metal electrodes, to which rf-voltage U is applied, angle θ is the Bragg's diffraction angle and 2θ is measured double-θ.

To clarify the origin of the non-structural light emission generated by a piezo-active acoustic vibration in MLNO, the measurements of NAIL intensity and spectra are made along with the acousto-electrical characteristics of the vibrating plates. In addition, the X-Ray diffraction by crystal surface, and PL properties of some electrically charged impurities are under study.

### 3. Results and Discussion

The amplitude characteristics of NAIL intensity $B$ are measured in a number of samples. The typical dependency is presented in Fig. 3.

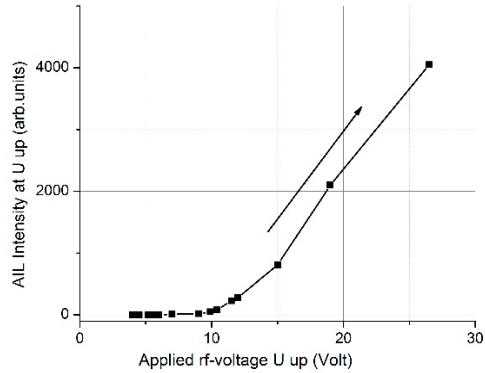

Fig. 3. Intensity of NAIL on excitation rf-voltage from MLNO sample LN6K1.

In Fig. 3, the voltage amplitude near 10 V corresponds to a threshold value for exciting NAIL. The measurement of an rf-electrical current $J$ through the sample shows that exactly at rf-voltage $U = 10$ Volts the current is characterized by a quadratic law of $J \sim U^2$, which testifies on the effect of electron injection from the metal contacts into adjacent crystal. The injected charge carriers must influence an rf-impedance of MLNO-sample, which was observed in our experiments as well. Thus, the strain of $10^{-5}$ is a threshold one for a number of nonclassical nonlinear effects in lithium niobate especially the NAIL. Experimentally it is also detected a generation of the Acoustic emission (AE) again under the over-threshold excitation voltage. The presence of AE at $\varepsilon > 10^{-5}$ means that the crystal lattice defects such as dislocations, their groups and others two-dimensional irregularities come to their motion exactly at $\varepsilon > 10^{-5}$. The vibrating dislocations and two-dimensional defects consisting of their groups such as micro-grain boundaries possess the electric fields that are not screened out any more by the charged point defects since the point and linear defects have very different dynamic properties. For instance, under over-threshold dynamic strain, the vibrating dislocations may leave the regions with charged point defects that compensate dislocation electric field. It happens simply due to the fact that the dislocations may vibrate with ultrasound frequency but the point defects mainly are sitting in their original sites within crystal lattice. This situation obviously may lead to an interaction between the charged defects and injected electrons.

Strikingly, taking into account such an interaction of injected electrons with charged defects can produce a simple tractable model of NAIL, which yields the excellent agreement with the experiment. The possible processes of energy gaining by the electrons and the subsequent light emission are sketched in Fig. 4.

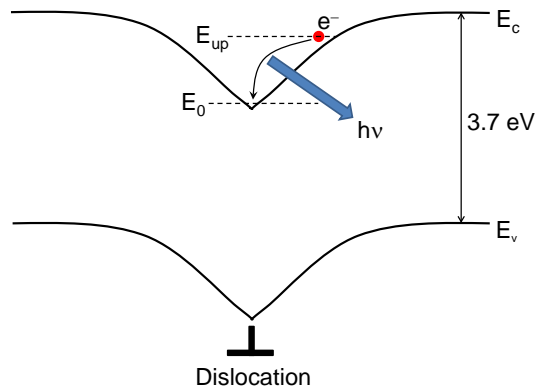

Fig. 4. Energy-band diagram of a LiNbO$_3$ with a dislocation in the center of the scheme. The dislocation is suggested to carry a positive charge.

The band in Fig. 4 is bending at the line defect, it is dominated by the dislocation charge. The light emission process with the energy hv arises in the electron (e⁻) transition shown by the tilted arrow. The optical absorption edge value of ≈3.7 eV is taken from Ref. [10]. The electric fields associated with dislocations may lead to an interaction between the vibrating charged defects and injected electrons, which lifts them up thus populating upper energy levels exemplified by the level $E_{up}$ in Fig. 4 in the conduction band $E_C$. It has been shown in a number of works that one can explain the broad electric-field affected luminescence either structured or blurred into the non-structured band by the superposition of different radiative transitions, which are resulting from transitions of electrons between different valleys in the bent conduction bands [10 - 15]. Assuming a Maxwell distribution of the carriers in the band bending valley, the emission spectrum produced by the electronic transitions to the bottom of the valley, which is shown by tilted arrow in Fig. 4, is given by the equation (1) below [11].

$$I \propto \exp\left(-\frac{h\nu}{kT}\right), \qquad (1)$$

where $T$ is the effective electron temperature and $k$ is the Boltzmann constant. It is found that the NAIL spectra are fitted to this equation very well, as exemplified in Fig. 5.

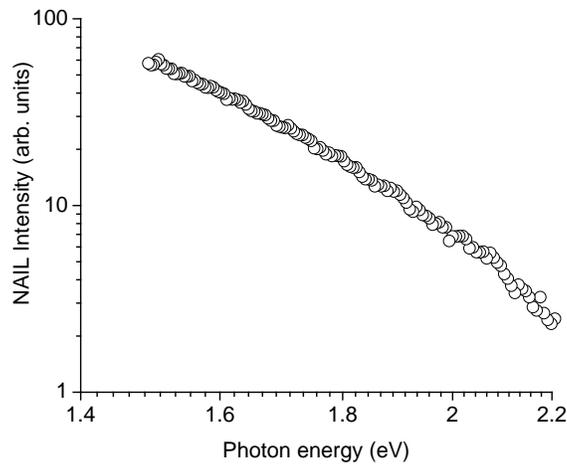

Fig. 5. Spectra of NAIL from MLNO sample. The fitting of the experimental points plot by the Eqn. (2) coincides with the experiment under the electron temperature $T = 2330$ K.

It is important to notice that the injected carrier heating in the dislocation electric field results in hot thermal distributions characterized by an electron temperature that can far exceed that of the lattice. Fitting the experimental NAIL spectra given in Fig. 5 to Eq. (2) yields the electron temperature $T$ of about 2330 K. Thus, one of the key requirements in developing model is the fact that the NAIL spectra are nearly linear on a log (NAIL intensity) – linear (photon energy) scale, as shown in Fig.5.

At micro-scale level, the acousto-dislocation interaction is independently verified with the help of X-Ray diffraction. The detailed description of that measurements are given in the reference [16]. This interaction may include the vibrations and motion of the dislocations along with their groups under the over-threshold acoustic deformations of $\varepsilon > 10^{-5}$. First, the Bragg angle $\theta_B$ is calculated from the experimentally measured double-angle $2\theta_B = 111.7$ degrees, Fig.2. Then the rocking curves are measured by a slight changing of an incident angle $\theta$ near the $\theta_B$. The rocking curves are measured without ultrasound and under the over-threshold acoustic amplitude corresponding to $\varepsilon > 10^{-5}$. The measured initial rocking curves without ultrasound and then under ultrasound are different. It is a consequence of the changes that acoustic wave produces to the micro-regions of a coherent X-Ray scattering. In other words, the boundaries of these micro-regions consisting mainly of the dislocations may dynamically change their locations in a crystal bulk, as described in details in ref. [16]. Thus the effect of acousto-dislocation interaction and their motion is obvious, and consequently the above described mechanism of NAIL radiation is plausible. The theoretical analysis of the rocking curves yields the possible dimensions of the blocks of coherent scattering of X-Rays in the range of (100± 20) microns; and in some cases it can be even higher up to hundreds of microns.

Since the point defects tend to have higher concentration near the dislocations and their groups, one can check the X-Ray estimate by measuring the impurity distribution in the samples. The strong photoluminescence (PL) of the F-center near 400.56 nm is presented in Fig. 6.

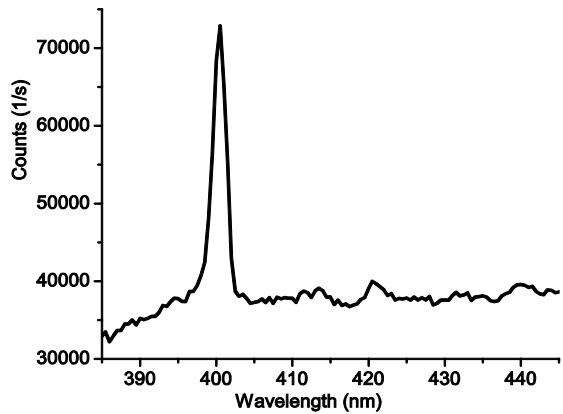

Fig. 6. Photoluminescence spectrum of F-center from MLNO-2K sample at an excitation wavelength of 310 nm.

It is interesting to note that the spectral positions of some lines in PL from MLNO practically coincide or are very close to lines in SAIL spectrum presented in Fig. 1, plot 3. Among them are Li, Nb, Ne, K, O, Fe and others. The distribution of the point defects along the wafer may be estimated with the help of PL spectra taken from different points along the wafer. This distribution turns to be not uniform. It has a quasi-periodical component of micro-scale spacing between the extremes positions along crystal from which the lines are obtained. The typical defect distribution along the Z-axis in MLNO is presented in Fig. 7. The excitation of luminescence is by an ultraviolet light.

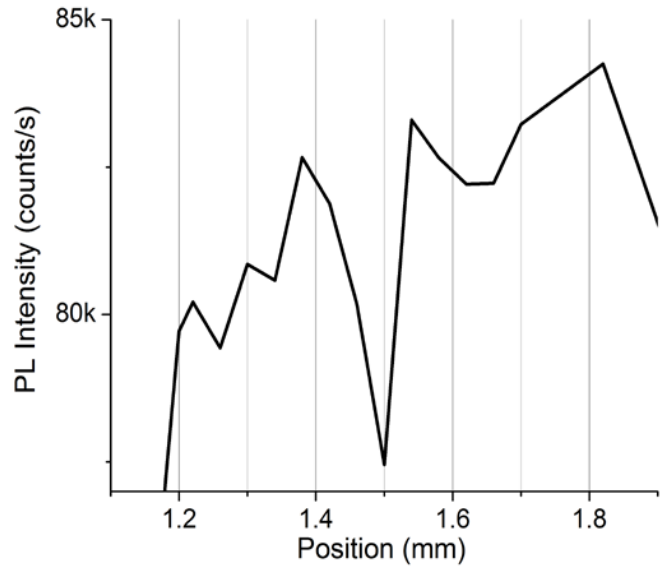

Fig. 7. Distribution of the F-center defect along the Z-axis in MLNO-2K sample.

## 4. Conclusion

In conclusion, the broad nonstructural acousto-injection luminescence in metalized LiNbO$_3$ is consistently explained by a model based on the interaction of injected electrons with charged defects including the dislocations and point defects. The charged point defects including those responsible for electrical conduction in MLNO tend to locate near the dislocations, which in turn are located in the vicinities of boundaries between the blocks of coherent scattering of X-Rays. The results of this research may be used for better understanding of acousto-electrical and acousto-optical phenomena in Lithium Niobate and for developing new solid state devices based on the ferroelectric ceramics.